\documentclass{iopjournal}
\usepackage{adjustbox} 
\usepackage[utf8]{inputenc}
\usepackage{xurl}

\begin{document}

\articletype{Paper} 

\title{How undergraduate physics students use generative AI for computational modeling}

\author{Karl Henrik Fredly$^1$\orcid{0009-0006-7788-1362}, Tor Ole B. Odden$^1$\orcid{0000-0003-1635-9491} and Benjamin M. Zwickl$^{1,2}$\orcid{0000-0002-8925-6912}}

\affil{$^1$Center for Computing in Science Education, University of Oslo, 0316 Oslo, Norway}

\affil{$^2$School of Physics and Astronomy, Rochester Institute of Technology, 84 Lomb Memorial Drive, Rochester, NY, 14607}

\email{karlhf@uio.no}

\begin{abstract} 
Generative artificial intelligence (genAI) is becoming increasingly prevalent and capable in physics, particularly for programming-related tasks. How, then, does genAI affect students’ computational modeling? We interviewed 19 undergraduate students who had recently completed an open-ended computational assignment that encouraged the use of genAI, asking them how they used it. We then conducted a thematic analysis of these interviews using a framework for computational modeling in physics. We found that genAI significantly impacts several aspects of students' computational modeling, such as the planning, implementing, and debugging of computational models. GenAI can also help students find resources and introduce them to new computational tools. Productive use of genAI was associated with students limiting its use to small steps in the modeling process and consistently double-checking the formulas, explanations, and code it provided. We also identified challenges students faced due to an over-reliance on genAI, such as working from false model assumptions and not spending time learning the fundamentals of computational modeling, especially debugging. Finally, we discuss implications for teaching, such as the need to teach students how to use genAI productively and to urge them to plan before they code. We also highlight the continued value of low-stakes assessment and teaching assistants for teaching computational modeling, as the task remains difficult even with the introduction of genAI.
\end{abstract}


\section{Introduction} 

The recent advancement of large language models has led to a substantial change in both physics education and research. In physics, like in many other fields, researchers can now use genAI tools to speed up routine tasks \cite{skogvoll_how_2025}, or in some cases even advanced workflows or discovery \cite{wei_ai_2025}. This ongoing development challenges both the content we teach and the ways we teach it. This development is especially pressing for computer programming, which genAI is especially useful for \cite{ai_index_steering_committee_ai_2025}. This introduces a twofold challenge to physics instructors, as they were already adapting to the introduction of programming in their physics courses, a change driven by the increased importance of programming for both physics research and jobs in industry \cite{noauthor_skills_2024, pold_field_2025, mulvey_physics_2025}. This challenge is now further amplified by the rapid adoption of genAI among students \cite{wang_scaffold_2025}, who are aware of this development and are already using genAI during their studies, with use cases as varied as giving explanations, finding sources, and cheating \cite{wang_scaffold_2025}.

In this ongoing development, there is a need to understand the challenges and opportunities these developments present for students’ learning of physics, especially in this meeting point between physics and programming. There are already ideas on how genAI can be useful or detrimental for learning, which we will cover shortly. However, there is still much we don’t know about how genAI affects students’ computational modeling and the role of instructors in guiding students' use of genAI. We therefore aim to answer the following research question:
\begin{itemize}
  \item \textit{When, why and how do students use generative AI for computational modeling when working on open-ended computational physics assignments?}
\end{itemize}
We will base our answer to this question on students’ own experiences, guided by what we know about genAI and computational modeling from education research.

\section{Generative AI in science education} 

The effect of genAI, specifically large language models (LLMs), in science education is still a relatively new area of research. However, there are already many ideas and findings on what LLMs can do, advantages to using them, disadvantages to using them, and what we should do about it.

Although the underlying architectures and procedures for training and using LLMs are known to us \cite{polverini_how_2024}, what LLMs can actually do is a constantly expanding target with no clear limit. The capabilities of these models are still increasing on a wide range of tasks, such as solving math or coding problems, understanding images, and general reasoning \cite{ai_index_steering_committee_ai_2025}. In physics, LLMs are already capable of solving a substantial number of analytical and computational problems at the undergraduate level \cite{pimbblet_can_2025}, even writing short physics essays just as well as students \cite{yeadon_evaluating_2024}.

In education, the use of genAI has been shown to have potential benefits. GenAI can guide students through problem solving \cite{tong_exploring_2025}, making them more efficient \cite{fan_beware_2025}. And genAI produced educational material, once quality controlled, has been shown to in some cases promote learning \cite{lu_incentivizing_2025} and student interest \cite{lademann_augmenting_2025} better than previously used materials.

Research has also identified several risks of genAI use in education. Although genAI can guide students through problem solving, students tend to use genAI as an answer giving tool rather than a collaborator, leading to worse performance than if they collaborated with others \cite{tong_exploring_2025}. The increased efficiency of using genAI might come at the cost of positive social dynamics \cite{tong_exploring_2025} and deep cognitive engagement, and the development of a dependence on the technology \cite{fan_beware_2025}. This dependence on genAI can be amplified by stress, and might lead to increased laziness, the spread of misinformation, and lower levels of creativity and critical thinking \cite{zhang_you_2024}. Even more worrying are the signs of lessened agency and longer term negative learning outcomes when using genAI \cite{kosmyna_your_2025}. In addition to these risks affecting students, genAI also threatens the displacement of learning and teaching assistants, whose development of educational competence is important for the entire educational system \cite{otero_physics_2010}.

To make the most of these possible advantages and mitigate these risks we need a better understanding of how students should use genAI in various contexts, and how to best promote these strategies. In science education specifically, there is a gap in the literature on the impact of genAI on scientific practices \cite{erduran_impact_2024} and implications for teaching. Of particular importance within this gap is the impact of genAI on computational modeling, where both programming and agency, elements which can be greatly impacted by genAI \cite{pimbblet_can_2025, kosmyna_your_2025}, play a central part \cite{phillips_physicality_2023}. The rich literature on this topic will guide our study.

\section{Computational Modeling in Physics} 

Computational modeling is a central practice in modern physics research, as well as in the application of physics to real world problems. Among physics PhDs, simulation and modeling are important skills, in addition to design, development, and programming \cite{noauthor_skills_2024}. In physics education, computational modeling offers students the tools to make complex physics problems tractable \cite{caballero_implementing_2012}, and serves as central a scientific practice that we wish for students to learn and engage in \cite{caballero_implementing_2012, irving_p_2017, burke_developing_2017, fredly_physics_2026}.

In this study, we focus on the practices students engage in when using genAI for computational modeling. For this purpose, we use the model of computational modeling in physics developed by Phillips et al. \cite{phillips_physicality_2023}, which was based on previous work by Burke and Atherton \cite{burke_developing_2017}. Their model has five components that together describe a non-linear cycle of \textit{production} and \textit{critique} practices that students engage in to reach certain \textit{objectives}. During this process, students use \textit{resources} to produce certain \textit{products} which can become resources for future inquiries. Figure \ref{fig:model} outlines our adaptation of this framework.

These five components are multi-faceted, such as the \textit{production} component, which includes descriptions of various practices like planning, implementing, interpreting and concluding. Or the \textit{resources} component, which describes various resources students can use, such as data, literature, and their own or others' expertise. The \textit{objectives} component describes both the students' epistemic objectives, their target knowledge or learning outcome, and the students' pragmatic objectives, such as wanting their code to be fast, or to get done quickly and get a good grade. The components of the model are interconnected. It describes how various resources are used in the production and critique of models, and how critique can both alter and be altered by the objectives of the inquiry. These five components make up their framework for knowledge production in computational modeling, which they applied to study an educational environment with a focus on computation, agency, and the construction of physical artifacts. The practices and connections are extensive, but were not meant to be comprehensive. Phillips et al. \cite{phillips_physicality_2023} anticipated adaptations of their model.

As our study has a different focus than the study by Phillips \cite{phillips_physicality_2023}, we adapt the framework to instead focus on how students use genAI for computational modeling (see Figure \ref{fig:model}). In the \textit{production} component of the framework, for instance, we describe students using genAI to plan, implement, optimize and interpret code, and the ways this use is different from students engaging in these practices without genAI. In the \textit{objectives} component, we focus more on the ways the students' objectives affect the ways they use genAI, and the ways genAI use might change both their epistemic and pragmatic objectives. We also include new elements which cover students' strategies for inspecting AI-generated code, and their use of genAI to help with physics theory and analysis. These adaptations constitute the results of our paper and will be covered in detail in the results section.

\section{Context} 

To study the effect of genAI on students’ computational modeling, we need a context where students engage in authentic computational modeling, a context where students produce their own models and code. The University of Oslo is one such context. The University of Oslo has incorporated programming into its physics courses for many years, a development which has been supported by significant efforts and resources \cite{odden_computational_2020, odden_using_2023, odden_using_2021}. In the physics bachelor's program, students learn programming in a dedicated course in their first semester, and go on to use programming in most physics courses, as well as many other non-physics courses. 

In this study, we interview physics students enrolled in a third semester course in electromagnetism. Students taking this course are typically in the middle of their physics-related bachelor's degree, and have taken at least one programming course, a mechanics course, and a calculus course. Some students take this course later in their studies and have even more experience with programming. The course includes weekly lectures and group work sessions designed to promote student participation and active learning. The course has two closed-book, in-person exams: one theoretical exam and one with an emphasis on programming. The course includes a number of mandatory assignments with both theoretical and computational exercises.

Additionally, the course includes a much larger mandatory assignment called a computational essay, which students complete and present alone or in pairs. A computational essay is a report of an inquiry into some computational problem. The essay combines elements of a typical report, like an introduction, theory section and conclusion, with the code used to compute the results. This combination of text, figures and code is made possible by computational notebook environments like Jupyter Notebook and Google Colab. Computational essays, including the ones in this course, have been the object of several previous studies, which have looked at their ability to foster creativity \cite{odden_computational_2020}, disciplinary epistemic agency \cite{odden_using_2023}, as well as their ability to scaffold professional physics practice \cite{odden_using_2021, caballero_teaching_2025}. The scope and open-ended nature of these assignments, as well as their computational component make them suitable for studying how students use genAI for computational modeling.

In this course, the computational essay assignment is open-ended, only requiring that students use computational models relevant to the course. The students are given a few example computational essays to take inspiration from or build on if they want, which many do, but are otherwise left to study whichever problem they want. Students were also required to hold a short presentation of their assignment in front of  a small group of peers and instructors. Though the assignment was mandatory, it did not count toward the students' final grade. Due to the lenient grading, low tuition fees, and low stigma around retaking courses, students were likely not overly worried about failing the assignment.

During the year data was collected, though genAI was not a big focus in the course, the students were given a short instruction on the use of genAI for the assignment \cite{malthe-sorenssen_computational_2024}:
\begin{quote}
    We encourage you to use ChatGPT or other language models both for developing text, code and for working with physics and mathematics. But remember that it is important to have a critical view of the input you get from the language model. If you use a language model, you should reflect on how the language model has influenced your work in a separate paragraph or where appropriate in the essay. Also think about how it is reasonable to reference the use of a language model and discuss it as part of your presentation. \textit{(translated)}
\end{quote}

\section{Method} 

\subsection{Data collection}
We recruited students for this study by visiting a lecture and work session in the third semester electromagnetism course, where we informed students about the study, obtained consent to collect their computational essays, and invited them to participate in interviews. We collected 40 computational essays, all of which were anonymized. We only used the essays from the interviewed groups for analysis.

The first author interviewed 13 groups, totaling 19 students. The student participants included 4 women and 15 men. There were 6 interviews with both group members, 4 interviewees worked alone, and 3 students were interviewed without their partner.

The interviews lasted around an hour and followed an interview protocol that covered the students’ background with physics, programming, and genAI use, before asking for a detailed account of their work on the computational essay. The interview protocol is included in the supplemental material. The interviewer read through the students’ essays in advance, and had themselves studied computational physics at the same institution some years prior, allowing them to ask pointed follow-up questions regarding the students’ choices and experiences. After the interviews, 7 groups provided genAI chat logs relevant to their computational essay work, and all interviewed students were given a gift card for their participation. We transcribed and anonymized all interviews, the names used in this paper are pseudonyms. The interviews were conducted during the fall of 2024. All 12 groups who used genAI used a version of ChatGPT-4o provided by the university, though three groups also used Claude, and some students used Perplexity and GitHub Copilot.

\subsection{Analysis}
In order to study the several ways in which students use genAI for computational modeling, we need a research method that will allow for open-ended analysis of many student experiences. We have therefore chosen thematic analysis of semi-structured retrospective interviews as our method. This method will allow us to extract rich and varied experiences from students, while still allowing analysis of a good number of students.

We analyzed the interviews using a 6-stage thematic analysis, as described by Braun and Clarke \cite{braun_using_2006}. First, during transcription, the first author noted down elements of interest and grouped them into initial themes, corresponding to the first two stages “familiarizing yourself with your data” and “generating initial codes”.

During the second stage, we leveraged the framework for computational modeling by Phillips et al. \cite{phillips_physicality_2023} to organize and interpret how and why students used genAI for computational modeling. We chose this framework because its structure and specific elements captured the breadth of our initial findings well, requiring only minor additions. Then, the first author formed these themes into an initial coding scheme and coded the data. During the coding, we refined the themes and their descriptions, as well as the arguments of this paper, following the third stage, “searching for themes”. After the coding came the fourth stage “reviewing themes”, where the first and second authors reviewed the themes and arguments by both coding approximately 15\% of the data. The first author gave the second author descriptions of the codes, two interview transcripts, and a document with the text segments that were coded by the first author. After the second author coded the segments we computed the Cohen's Kappa statistic to measure agreement, which gave a value of 0.62, indicating moderate agreement \cite{mchugh_interrater_2012}. After this, the first and second author discussed any differences and came to an agreement on all segments. We consider the agreement to be solid, as most differences in the initial coding were due to overlapping perspectives inherent to the Phillips computational modeling framework, rather than opposing interpretations of the data. Finally, we modified the themes based on the review and in the process of writing this paper, which constitutes the final two stages “defining and naming themes” and “producing the report”.

In addition to this analysis of the interviews, we used the computational essays and genAI chat logs to triangulate different themes, such as how students documented and optimized their code, or how they used genAI for code debugging and explanations of the physics.

\section{Results: How students use Generative AI for computational modeling} 

\begin{figure*}[t]
    \begin{adjustbox}{width=1.4\textwidth,center} 
        \includegraphics[width=\linewidth]{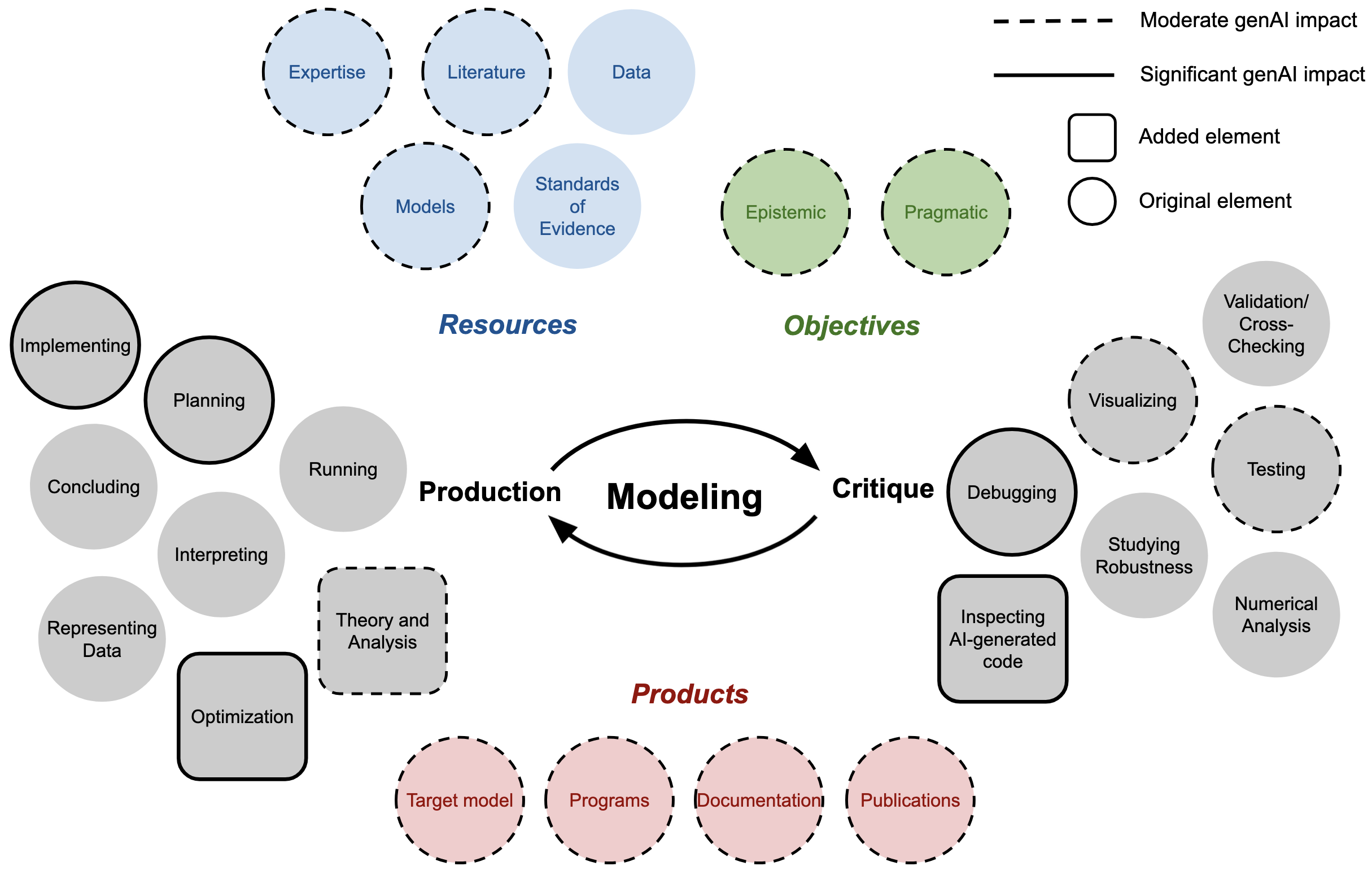} 
    \end{adjustbox}
    \caption{How students' use of generative AI impacts the different elements of computational modeling. Adapted from Figure 2 of Phillips et al. \cite{phillips_physicality_2023}. In a modeling inquiry, students iteratively cycle between \textit{production} and \textit{critique} practices, some of which are significantly impacted by genAI, as students either commonly used genAI when engaging in these practices, or performed the practice in a substantially different way than if they did not use genAI. Students engage in these practices to reach certain \textit{objectives}, and use \textit{resources} and produce \textit{products} during the process, some of which are affected by genAI in moderately impactful ways.}
    \label{fig:model}
\end{figure*}

We organize our results according to the five components of computational modeling described by Phillips et al. \cite{phillips_physicality_2023}, as these components capture the breadth and interconnectedness of the different practices and elements of computational modeling. Building on this model, we describe how \textit{some} students used genAI for computational modeling, why they chose to use it, and the degree to which genAI changed what students were doing. These findings are summarized in Figure \ref{fig:model}, which borrows the structure of Figure 2 in the Phillips et al. paper, though it has been simplified by not including the arrows signifying connections in the original model.

Our main addition to the model is our description of how much genAI affects the different elements of computational modeling. We found some elements to be significantly affected, specifically practices within the production and critique categories. These were the planning, implementing, debugging and optimization of code, as well as the inspection of AI-generated code, a new element we added to the model. We denote these practices as significantly affected by genAI, as students either commonly used genAI when engaging in these practices, or performed the practice in a substantially different way than if they did not use genAI.

Other elements were only moderately affected by genAI, such as the students' work with physics theory and analysis relevant to the model, the expertise and literature students relied on, and the products of the modeling process. We denote these elements as moderately affected, as the students' use of genAI did not substantially change what they were doing, as the genAI use was rare and only moderately impactful. Finally, we found some elements to not be affected in any meaningful way, like the students' strategies for running their code, validating their results or collecting data. 

Looking at genAI use more broadly, 8 of the 19 interviewed students used genAI frequently to assist with finding and explaining information and with coding. Another 9 students used genAI more sparingly or reluctantly, either using it for a smaller subset of tasks, or when they had been stuck for some time. One student tried using genAI for a few tasks, but found it unhelpful and unwanted, and one student was completely against using it. These frequencies do not represent the student population at large, as students who used genAI were probably more open to participating in a study about their genAI use. In what follows, we will describe the different ways students used genAI in more detail.


\subsection{Production} 

The first component of computational modeling we consider is production, which covers students' production of models, code, theories and documents to iteratively be critiqued and refined, before finally being published or used as a resource for a new modeling inquiry.

\subsubsection{Planning} 
Students generally started by coming up with an idea for their essay on their own, or by choosing an example essay to work from. Students had little trouble with ideation, and preferred coming up with ideas on their own rather than using genAI. After their initial idea of what phenomena to model was formed however, students often had more trouble. They found it difficult to apply concepts from electromagnetism numerically at an appropriate level. Many students turned to genAI at this stage. Some asked it very broad questions about the thing they wanted to model, others asked more pointed questions about how to model their chosen phenomenon in their specific context. Students found genAI used in this way to be helpful for getting them started, and for adding or removing elements to reach the right level of complexity.
\begin{quote}
    \textbf{Levi: } That’s maybe one time where I got something out of it. That it became like, okay, now I have a starting point. It turns out that what ChatGPT suggested was wrong. But it didn't matter that much then. Because you get a starting point doesn't need to be right. As long as it makes me think of something that later gives good ideas.
\end{quote}
However, students who relied too heavily on genAI at this stage had worse results. GenAI often made incorrect assumptions, and without careful consideration along the way students risked wasting time or ending up with results they did not understand
\begin{quote}
    \textbf{Ivar:} I understood the basics of the problem and I had coded from the book what we learned, like the Laplace equation, von Neumann technique. So I'd learned these and I'd coded them and then I would use them with ChatGPT and then I would make the problem more complicated. But I reached such a complicated far point that I couldn't even explain the code that I had now created.
\end{quote}
In addition to helping with planning out the assumptions, formulas, and methods to use, students also used genAI to help plan how to write the essay and code. Some students used genAI to provide an outline based on the assignment text and their own initial ideas, to help get them started.
\begin{quote}
    \textbf{Christian:} I know at least that I used ChatGPT pretty much straight away, just to get some kind of idea, some kind of sketch or something. Because I thought it was really good to just have some kind of template to write from. And then I just gave it to him like that, ``I was going to make a computational essay, this is the topic [survivability of a fish in a microwave]".
\end{quote}
The students' experiences with using genAI for planning were very mixed. Some found it very helpful, some felt they wasted time correcting the genAI's mistakes, while others only used genAI once they got stuck, preferring to try implementing a model on their own first. The students did not reflect much on how outsourcing this important part of modeling might affect their learning.

\subsubsection{Implementing}
With an initial idea of what to model and how to model it, students began implementing the model in code. The nine students who felt that they got the most out of using genAI for coding all used it in a similar way. These students used genAI to help implement specific, well-understood parts of the model. Examples of such small parts could be finding a value in a list that was closest to some other value, rewriting code to use loops, adding a dimension or complicating factor to an already working simulation, or producing a specific type of plot from simulated data.


A few students offloaded even bigger and more complex coding tasks to genAI, like translating over 1000 lines of C++ to Python, or setting up a 3D visualization. Even though these students quickly identified mistakes in this AI-generated code, they found it helpful to use this wrong code as a starting point for progressing in the assignment.
\begin{quote}
    \textbf{Alec:} For instance, in the computational essay, I was wondering a little bit about how to simulate the magnetic field from a ring. Because I had done it before just with a charge density in a line, like I did in physics, but I was wondering a little bit about how to do it in a ring. So I asked it[ChatGPT]. It gave me the code for charge density in a ring, but it was with electric fields, so it's a bit of the same. So I found out that it gave me a bit of the wrong code, but it gave me an insight into how to do it.
\end{quote}
Offloading programming to the genAI has some risks however. If the student doesn't understand the code generated by the genAI, they are unlikely to find its mistakes, might learn the wrong things, and will be unable to produce a good assignment. Also, genAI is unlikely to push back against students' faulty ideas for what to implement, and will often encourage any approach the students suggest. This was the case for Ivar earlier, who wanted to model a speaker, but did not understand the assumptions underlying the code, and was not able to effectively find mistakes introduced by the genAI. Here he is reflecting on his strategy of generating code first and working on understanding the equations later:
\begin{quote}
    \textbf{Ivar:} I fought with the equation way too late. Does it make sense? Find the equation that you want to use in the code first. Because that's what I'm learning now. You should just really go to the equation that you want to use in the code, then code. And not the opposite. Don't code... And then find out what equation is really good for this code.
\end{quote}

\subsubsection{Optimization}
\begin{figure*}[t]
    \centering
    \includegraphics[width=250px]{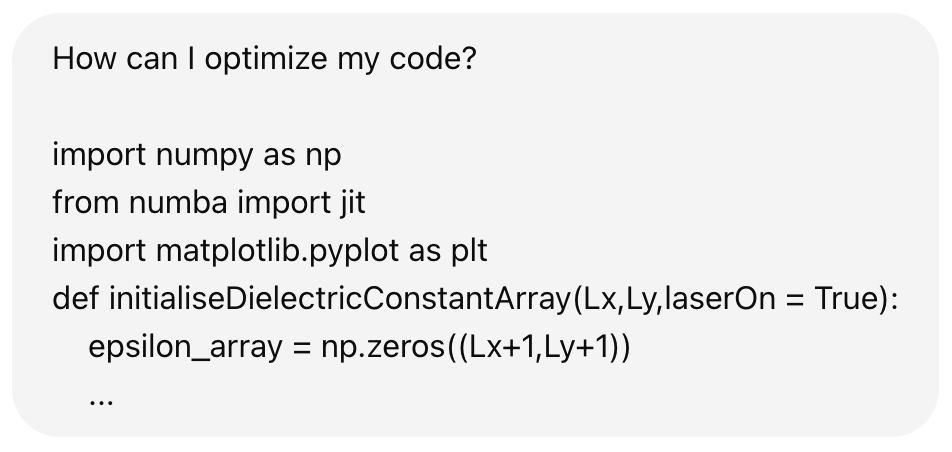}
    \caption{The first part of the prompt a group of students sent to ChatGPT to optimize the Poisson solver provided by the course.}
    \label{fig:prompt_opt}
\end{figure*}

In addition to using genAI to get their code working, two groups used genAI to make their code faster. One of these groups wrote an essay on simulating lightning, and based their code on a Poisson solver provided by the course. Unbeknownst to them, the solver was implemented inefficiently on purpose, so that students would have to figure out a way to optimize it. When they found the solver to be too slow for their needs, they pasted the entire code into ChatGPT, as illustrated in Figure \ref{fig:prompt_opt}, and were provided with several suggestions they could incorporate into a much faster code, though with a few alterations.

By using genAI, these students were able to vectorize or otherwise speed up their code, even though optimization often requires more experience and niche knowledge than typical third semester students have. And though the genAI did most of the work in finding the optimizations, the students felt that it was helpful for their learning:
\begin{quote}
    \textbf{Samuel:} And then I thought that as long as it doesn't write the code for us, it's fine to have it optimize it a little, in a way. Then we can learn how to optimize that type of code, I thought.
\end{quote}
Using genAI to help optimize code not only made the students' code faster, but also allowed them to see optimization tricks they normally would not encounter. With repeated exposure to such tricks, or a deeper dive into how they work and when they are applicable, motivated students might then learn how to optimize code they encounter later.

\subsubsection{Theory and Analysis}
Although the students had mixed opinions on how genAI should be used for programming, all students were skeptical of using genAI to solve problems they had with physics theory or equations. The students argued that genAI simply performed worse at physics than at programming, based on their experience using it and the lack of genAI training material for their context. One student also highlighted the value of working through the analytical part of physics on his own:
\begin{quote}
   \textbf{Aaron:} I feel like... that’s just how I like to work with physics. I like to try to go through the math formulas myself and try to understand them on my own before I go to Google or anything like that. So, that’s what I do.
\end{quote}
When some students did use genAI for theory and analysis, it was typically for help with explaining theories or manipulating equations they felt they could easily double check with their underlying understanding or external sources. Some students particularly valued genAI's help with linear algebra.

\subsection{Critique}

The second component of computational modeling we consider is critique. Here, we cover both the ways students used genAI to critique their own work, and the ways students critiqued or did not critique genAI output. We exclude some aspects of critique from the original model, as the students in this study rarely used genAI for validating their models with different calculations, and they rarely studied the robustness and the numerical performance of their models.

\subsubsection{Debugging} 
Using genAI to debug code was the most common use-case of genAI among all 12 groups who used it. When the students encountered an error message or incorrect result, around half took the problem straight to genAI. In many of these cases, debugging was as simple as the student pasting their code into the chatbot and asking it to fix it.
\begin{quote}
    \textbf{Christian:} It's really, to start with, just put everything into ChatGPT, just like, ``Fix this", and then just, usually it actually works, to be completely honest.
\end{quote}
Students found debugging with genAI to be very useful, and were often surprised at how quickly it solved issues they had been working to fix for several minutes or sometimes hours. Debugging these kinds of models can be quite time-consuming, even with the help of an instructor, so students appreciated the efficiency of debugging with genAI. However, this efficiency also meant that students did not take the time to learn how to identify errors in their program, a trade-off several students noted as hard to manage.

The genAI sometimes got things wrong while debugging. Often the bug remained, or the genAI introduced new bugs which could frustrate students and waste their time. When the code had a physics error, like in the choice of method or physical assumptions, students found genAI to be particularly unhelpful, as it often did not resolve these underlying issues. Instead, the genAI often egged the students on with new additions and ``fixes" to the model, like this case with Vegard, who was simulating the fields inside a tokamak reactor:
\begin{quote}
    \textbf{Vegard:} I kind of brute-forced a little too much to just make the code work by asking ChatGPT why the code doesn't work. But when we found out what the problem was, it was actually a pretty simple problem, or it was just a problem that we forgot to include a field, a type of field, that was on Wikipedia for example or on other shorter basic article websites. So yeah, maybe I went a little too fast to just ask ChatGPT what the problem was.
\end{quote}
In some cases, however, the genAI provided parts of a solution, even though the code still did not work, which students could incorporate to implement a proper solution.
\begin{quote}
    \textbf{Alec:} Often it tries to fix it, but it doesn't fix it properly. So I often have to do it myself, but it gives me an insight into what's wrong.
\end{quote}
As the genAI was limited in what it was able to fix, the most productive students used genAI to fix small, concrete issues where they had some understanding of the problem and a sense of what a solution should look like. Though as the examples show, many students used AI to take substantial shortcuts when debugging. In our case, 12 of the 19 students took such debugging shortcuts with genAI.

\subsubsection{Validation, Testing and Visualization}
In order to verify the correctness of their code, most students simply ran it and checked the result with known values from theory, online sources, or their intuition. For some students, this type of check was sufficient for accepting their code and model, even if they didn't fully understand or trust the code they or a genAI wrote.
\begin{quote}
    \textbf{Henning: }But I still don't understand what I did wrong, and what ChatGPT did differently, because it looked pretty much the same. But it worked, and I had worked so much on it, so I just accepted it.
\end{quote}
One student, Ivar, formalized and automated these checks by making explicit tests in their code. Writing explicit test functions is not common in introductory courses, though this student had learned the basics in a programming course they had taken. This student used ChatGPT for writing his tests and was pleased with their implementation. Though as earlier quotes have illustrated, Ivar also offloaded a significant portion of the modeling process to genAI, leading to an overall inefficient process and incorrect results.

Students very often used genAI to write the code that generates plots, as customizing even simple plots can require many lines of code, requiring students to remember several plotting functions and their specific arguments. In these cases, students generally found it simpler to tell the genAI the title, colors, labels and line-styles they wanted, instead of looking up specific documentation for each setting. A few students saw such tasks as irrelevant for their learning, as they take time but offer little learning benefit:

\begin{quote}
    \textbf{Christian: }I personally think that the best thing about ChatGPT, and what I use it for the most, is not having to do things that I find boring. Like writing plots, or coding things that I know I can do.
\end{quote}

\subsubsection{Inspecting AI-generated code}
Whereas previously, students either wrote code from scratch, or adapted code examples from textbooks or online forums, students can now be given code by genAI which is already adapted to their specific problem. What is left for students to do when the code seems to work?

All students who generated code with genAI inspected it to make sure it was correct, though many students were not particularly thorough. The most common approach to inspecting AI-generated code, in addition to the aforementioned running, testing and visualizing, was seeing if the code ``looked right". Several students described their approach as quickly looking over AI-generated code to see if it had the elements they expected it to have, instead of parsing the logic closely.
\begin{quote}
    \textbf{Alec:} It's like I expect, like often when I ask, I expect to have something, like I know something about it. So I know what it should look like, that answer. So I can often quickly see if it's not quite right, based on how it looks, or how the code looks.
\end{quote}
These students were pleased with this superficial approach to inspecting AI-generated code, and stressed the importance of roughly knowing what the code should look like and do, to keep the scope of genAI questions within what they are able to verify.

When students were unsure if the AI-generated code ``looked right", some used genAI to provide explanations of the different parts of the code. Others, like mentioned earlier, rewrote the AI-generated code in a way that they better understood. A handful of students however, admitted to using AI-generated code they did not fully understand. One example was George, who used genAI to translate C++ code he found online to Python, even though the code was quite complex, and he only knew a little C++ from a course he was taking:
\begin{quote}
    \textbf{George: } The thing is that I know enough C++ that I can see the differences and recognize them. So I looked over what it had done, and I understood most of it, but then parts of the article went into frequency space and something like that, or some Fourier analysis and stuff, which I have no control over. So it became... Then I just had to trust that the AI did it correct there. Because I have no idea what is done there, either analytically or numerically.
\end{quote}
The students who inspected their code closely also did not use genAI to generate substantial amounts of code. These students preferred to stay in control during both the production and critique of their model.

\subsection{Resources}
The third component of computational modeling we consider is the collection of resources used. Here, we cover how students used genAI to supplement or replace the expertise and literature used when modeling, as well as how they used genAI to select an appropriate model.

Some students preferred using genAI instead of asking for help from instructors or teaching assistants. These students appreciated how the genAI was always available, and were sometimes hesitant to pester teaching assistants with constant questions or with questions the students considered too simple. One student also argued that genAI is much more efficient at helping with coding problems than teaching assistants, as they don't really have time to read through, understand, and untangle all of their code. For other queries however, most students still found the teaching assistants and instructors very helpful when they were available, especially for validating their models.

Many students used genAI to more efficiently find and understand online resources. These students viewed genAI as better at finding the information they needed than doing web searches or parsing course material. The students often preferred the explanations given by genAI, as they were more tailored for their specific needs than material found online or in textbooks, especially when their problems went outside course curricula.
\begin{quote}
    \textbf{Henning: }Maybe accept that... AI is better at teaching than online resources. It is much better at finding exactly what you are looking for than searching for it on Google. On Google you only find specific things, while ChatGPT can generate something that is tailored to your question. Instead of you finding the closest specific thing that already exists on the internet with Google.
\end{quote}
This ability of the genAI to tailor a response to the students' specific needs was especially appreciated by students when it came to the choice of model. Like we discussed on the practice of planning, some students found it especially difficult to find an appropriate model for the phenomena they wanted to study at the correct level of complexity. For these students, genAI played an important part in helping them manage the complexity of their models.

Even though most students found genAI helpful for finding information, they all later verified the information with trusted sources such as instructors, textbooks, or quality online resources, as they had little confidence in factual information provided by genAI.
\begin{quote}
    \textbf{Ane:} I used it in chemistry this year too. And there was just a lot of what he(the genAI) said that didn't make sense at all. And then I asked the teacher, and they were like, yeah, that's wrong.
\end{quote}
Some students did not use genAI for finding or explaining information, or only used it as a last resort. They preferred learning new concepts using the textbook, which they saw as more reliable and less prone to over-complicating the material.

\subsection{Products}

The fourth component of computational modeling we consider is the set of products students make while modeling. In this study, the computational essay is the main product. The essay has several "sub products", such as text, code, and figures, which were affected differently by genAI. Most students said that they did not use genAI for writing the text of their essay, though a few said they used genAI to improve their grammar, or to help with writing equations with LaTeX. Only one student reported using genAI to generate the entire text of their essay. This student felt pressed for time and made their essay by piecing together AI-generated explanations of his AI-generated code. The end result had goals, equations, code and results, but had several errors and lacked cohesion. The student said that he ashamedly admitted his strategy to the teaching assistant, who nonetheless accepted the essay and his presentation.

Students more commonly used genAI for producing code. As we described in more detail in the Production and Planning sections, many students used genAI to implement their model or fix their code, often by directly copying AI-generated code, or by rewriting certain parts of the genAI suggestions in their own style. A few students also used genAI to add comments to their already finished code. Students felt that the production of graphs was especially suitable for genAI assistance, and one group even used genAI to generate an illustrative image for their essay. 

Students using genAI also accumulated a collection of AI chat logs which could be used to review material. Students found it important to manage the information contained within a genAI chat ``session" in order for the AI chatbot to perform well. Students often started new chats when the genAI veered off course, or copy-pasted information from one chat into another to get better answers. In this way, students were not only producing code and notes for their essay, but also managed the context which the genAI used to produce its responses.

\subsection{Objectives}

The final component of computational modeling we consider is the set of objectives of the students' work. Here, we cover ways students' objectives affected their use of genAI, namely how they balanced their learning outcomes with the efficiency and quality of their modeling. We also look at how students changed their objectives as a result of their genAI use.

\subsubsection{Epistemic Objectives}

Every student we interviewed highlighted the importance of understanding the arguments, equations and code produced during their project. For these students, learning the material was clearly an important objective. The fact that the assignment was obligatory, but not graded might have supported learning as an objective.

Because of this focus on learning, most students were mindful of the ways they used genAI. They felt that if they outsourced the solution to genAI instead of solving the assignment themselves, they would not learn anything. Students therefore tried moderating their genAI use, so that it would improve their learning, or at least not hinder it.

For some students, understanding genAI suggestions was enough to feel like they were still learning what they needed.
\begin{quote}
    \textbf{Levi: }As long as I quality-assured what AI comes up with, and understand what I submit, I think it's just another tool. So I don't see any fundamental difference between artificial intelligence and a textbook as long as you understand that both are just a starting point for creating something of your own.
\end{quote}
Other students however, preferred trying to solve problems on their own. These students were worried about becoming too reliant on genAI if they used it too early in the problem-solving process, and only used genAI reluctantly when they needed it to finish the project.

\begin{quote}
    \textbf{Amanda: }I use it as little as possible, but as much as necessary.
\end{quote}

Some students also moderated their genAI use to promote learning by using genAI mainly to explain code and concepts, instead of using it to solve the problems directly. Several students felt that genAI used in this way helped them learn the topics they wanted to explore. 

\subsubsection{Pragmatic Objectives}

Although the students valued their learning while working on their essays, students also noted other factors which influenced how they used genAI.

Around half the students noted time-pressure as an important influence on how they used genAI. For these students genAI was seen as a helpful tool for saving time, especially when it came to writing code. Although students had around 3 weeks to finish the essay, and most students reported only needing a couple days to finish, many felt pressed for time due to postponing the work or needing to spend time on other tasks as well. For one student, George, this time pressure, together with the challenge of writing complex code, led to them using genAI for coding for the first time.

\begin{quote}
    \textbf{George:} I haven't really used AI to program anything until now in the computational essay. [...] Since it was such a large and complex project, and there was a bit of a time crunch. So then it had to be a bit more efficient. And then it was more efficient to have it generate a first draft, and then work on it myself.
\end{quote}
Some students who used genAI due to time pressure felt that they lost out on a learning opportunity, or lost out on the enjoyment of working on the problem on their own. These students seemed reluctant to use genAI, but still did, indicating that it is not enough for students to understand the risks of using genAI in order to stop them from using it in ways that are bad for their learning.

One of these students who felt that they offloaded too much of their work to genAI noted that he increased the scope of his project in order to feel like he put in the right amount of effort.
\begin{quote}
    \textbf{Nils:} A big result of using ChatGPT I feel is that our task is more extensive. We ended up doing more, because we had to feel like we had done something, in a way.
\end{quote}
This group was unique in how they alleviated the loss of learning when using genAI, as most other groups who used genAI instead tried to limit their use to what they deemed most necessary and useful.

\section{Discussion} 

\subsection{Summary of findings}
Returning to our research question, ``\textit{When, why and how do students use generative AI for computational modeling when working on open-ended computational physics assignments?}," we have found that students mainly use genAI for writing and debugging code, either when they are unsure of how to implement a model they have in mind, or when their code does not work. Students also used genAI to explain coding and physics concepts, or to find sources, but did not trust its suggestions without confirmation from their own expertise, trusted sources, or teaching staff. This distrust had its limits however, as several students used AI-generated code or models they did not understand in their final submission.

The students mainly saw genAI as a way to save time, and had mixed feelings on how it impacted their learning, with some students feeling like they were able to learn and do more, and some students feeling like they harmed their learning by using it. Despite these risks, students generally found genAI useful, though at times challenging to use effectively. And finally, even with the help of genAI, students still found computational modeling challenging.

\subsection{Implications}
Based on these findings, we suggest several implications for teaching. First, teachers and teaching assistants remain vital to the learning environment, and should continue to be present and available. Even students who used genAI extensively appreciated the availability and usefulness of the teaching staff, though mostly for conceptual questions about the physics of their models, rather than for programming-related problems. Additionally, teaching staff play a key role in verifying students' physics and programming knowledge while allowing them to make their own decisions.

Second, instructors need to lead students toward using genAI in a way that ensures they understand and verify the results along the way. While some students felt very comfortable with their strategy and results when using genAI, several students encountered problems with being led astray or not getting the learning experience or outcome they wanted. If students are given little guidance on how to navigate these issues, while at the same time struggling with both the content and time-pressure, many students might end up with poor learning outcomes.

Finally, open-ended assignments with genAI-access are unlikely to help students develop foundational skills in programming and modeling. Although most students in this study tried moderating their use of genAI in order to learn more, many spent only a handful of minutes trying on their own before turning to genAI to get unstuck. Although students using genAI in this way might be more efficient in the short term, their long-term learning is likely negatively affected, as they do not learn to solve modeling and programming problems on their own. Other forms of instruction and assessment which restrict students' access to genAI are necessary to ensure students develop core modeling skills.

Comparing our findings to prior research, we see that students' varied use of genAI aligns well with how physics professors use genAI. Physics professors also use genAI for planning, debugging and improving code, as well as for writing simple scripts, like plotting code \cite{skogvoll_how_2025}. Also, both students and instructors are distrusting of genAI answers, which might indicate that students are using genAI in a way that is authentic to professional physics practice, as the students' efforts to verify and critique genAI answers align well with the practices and wishes of physics professors \cite{skogvoll_how_2025}. However, physics professors already have a strong background in coding and computational modeling, and are much better positioned to evaluate the correctness and quality of AI-generated code than students who are new to coding. As noted by Shen et al. \cite{shen_how_2026}, genAI might increase the efficiency of professionals, but lead to decreased skill acquisition for novices. In this view, which we share, students should not use genAI like professors, but instead focus on developing fundamental skills.

Our findings also align with risks of genAI use identified by prior research. Students often used genAI because of stress \cite{zhang_you_2024}, and used it more as an answer-making machine than as a collaborator \cite{tong_exploring_2025}.  Furthermore, students often asked genAI for help instead of peers or instructors, which could undermine the collaborative nature of computational modeling \cite{tasar_physics_2023} and reduce the impact of instructors in helping students learn \cite{otero_physics_2010}.

GenAI impacts students' computational modeling both significantly and unevenly, and we must therefore reconsider which practices are integral to students' computational modeling. In particular, we should acknowledge the possibility of students outsourcing substantial parts of computational modeling to genAI, as was the case in how students in this study used genAI to debug code. We should also identify computational modeling practices that \textit{should} be partially delegated to genAI, where students and professionals can save time with little downside, like finding relevant sources or configuring simple plotting code. These uses of genAI, which can both harm and promote learning, challenge our understanding of what it means for students to do computational modeling.

\subsection{Limitations and future work} 
This study has several limitations. Firstly, our findings are based on retrospective interviews, which give limited insights into the moment-to-moment decision-making and strategies of students when using genAI for their computational essay. Secondly, we have limited data on how the use of genAI might influence the dynamics of the group or classroom. Thirdly, the students recruited for this study were aware that the interviews would be about their use of genAI. We might therefore have a skewed perception of how common different strategies for using genAI might be, as students who either struggled with using genAI, or heavily used it to complete the assignment for them might be less likely to want to talk about it. And finally, many of the students interviewed had little experience with using genAI. Students with more experience with genAI might develop more habits of misuse, or alternatively find new ways that genAI can help them learn and save time without being led astray.

Future work on computational modeling with genAI should look more into how the use of genAI affects student learning, their feeling of ownership and mastery, and the quality of their products. While we have given an overview of different ways students used genAI for computational modeling, the outcomes of this use are still not well understood, and are likely to change as the technology changes in the coming years. There is also need for more research on the use of genAI in less open-ended assignments and on graded assignments, where students might feel less of an incentive to experiment and explore the topic. Finally, future work should explore how genAI is being adopted by professionals, and whether and how students should prepare for this workflow. Specifically, if students should learn to effectively inspect, validate and build on AI-generated code, or if foundational computational modeling skills should be prioritized, which for now seems to be the case.

\section{Conclusion} 

We have found that students can use generative AI in many different ways to assist them in computational modeling. Especially when it comes to writing and debugging code, we see that genAI can allow students to build more complex models and save time. However, these advantages come with significant risks to student learning, which we saw with many students outsourcing real modeling and coding problems to genAI, and sometimes not even understanding the code or models they were building. 

However, we also see many students moderating their use of genAI to improve both their learning and the robustness of their results. This mindful use of genAI was in part driven by the students' ownership of the modeling and learning process, which was encouraged by lenient grading, upcoming coding exams, and the open-ended nature of their assignment. We also found that students did not rely on genAI for validating information when they felt that other sources of assistance, such as teaching assistants and textbooks, were useful and available.

Based on these findings, we are cautiously optimistic that genAI can be a helpful tool for enabling students to develop computational models that align with their interests. It can be a tool for getting students unstuck and helping them learn both programming and physics concepts, if used in the right way. However, instruction and assessment without genAI still play a vital role in teaching students the fundamentals necessary to validate their models and theoretical understanding.

This look into how students use genAI for computational modeling shows promise for how students might learn and choose to use genAI in helpful ways for learning and doing science, but more research is needed to understand the learning outcomes of students using genAI in different ways, and what instructors can do to promote genAI use that supports both productivity and learning.

\ack{}
This work was funded by the Norwegian Agency for International Cooperation and Quality Enhancement in Higher Education (DIKU) which supports the Center for Computing in Science Education.

\bibliographystyle{vancouver}
\bibliography{export-data}

\end{document}